\documentclass[
reprint,
nofootinbib,
 amsmath,amssymb,
 aps,
]{revtex4-2}

\usepackage{graphicx}% Include figure files
\usepackage{dcolumn}% Align table columns on decimal point
\usepackage{bm}% bold math
\usepackage[colorlinks = true,
        urlcolor  = blue]{hyperref}
\usepackage{appendix}
\usepackage{upgreek}

\usepackage[caption=false]{subfig}
\usepackage{tikz}
\usetikzlibrary{positioning}

\begin{document}

% \preprint{APS/123-QED}

\title{Demonstration of a novel phase space painting method in a coupled lattice to mitigate space charge in high-intensity hadron beams
}

\author{Nicholas J. Evans\textsuperscript{$\ddagger$}}
\email{nhe@ornl.gov}
\author{Austin Hoover\textsuperscript{$\ddagger$}}
\email{hooveram@ornl.gov}
\author{Timofey Gorlov}
\author{Vasiliy Morozov}

\affiliation{Oak Ridge National Laboratory, Oak Ridge, Tennessee 37830, USA}

\date{\today}

\begin{abstract}

Multi-turn charge-exchange injection is the primary method of creating high-intensity hadron beams in circular accelerators, and phase space painting during injection enables tailoring of the accumulated phase space distribution. A technique we call \textit{eigenpainting} allows injection of particles into a single mode of a coupled ring, providing full four-dimensional control of the phase space distribution. Under ideal conditions, uniform eigenpainting generates a linear-force equilibrium distribution in the transverse plane, with zero volume in four-dimensional transverse phase space, even including space charge. We have implemented eigenpainting for the first time in the Spallation Neutron Source (SNS) Accumulator Ring. Injecting 8.8~$\mu$C of 800~MeV beam, we obtain a final ratio of intrinsic transverse emittances of $\approx$2.4. We analyze the effect of space charge on the final distribution through comparison of the reconstructed phase space to particle-in-cell simulations.
\end{abstract}

\maketitle

\def\thefootnote{$\ddagger$}\footnotetext{These authors contributed equally to this work.}\def\thefootnote{\arabic{footnote}}

\textit{Introduction}---In high-intensity, low-energy hadron accelerators, space charge forces are a major driver of halo formation and beam loss \cite{cousineau_high_2015}. Losses can be driven by a variety of mechanisms, such as periodic crossing of resonance stop-bands \cite{franchetti_resonance_2010}, particle-core interactions \cite{holmes_dynamics_1999, liu_structure_2024}, and coherent instabilities \cite{hofmann_anisotropic_1998, oeftiger_dynamics_2021}. Several techniques have recently been proposed to mitigate space charge effects, including circular mode optics~\cite{burov_circular_2013}, pulsed electron lenses \cite{oeftiger_pulsed_2024} and nonlinear integrable optics \cite{danilov_nio_2010}. Here, we focus on controlling space charge effects by shaping the beam distribution as it is formed over hundreds or thousands of turns via charge-exchange injection in an accumulator ring. In facilities such as the Spallation Neutron Source (SNS) \cite{henderson_sns_2014} and the Japan Proton Accelerator Research Complex (J-PARC) \cite{Yoshimoto:2006wv}, time-dependent dipole (kicker) magnets modify the location and angle of the closed orbit during injection, updating the phase space coordinates of the injected beam centroid as a function of time. This process is known as \textit{phase space painting}. While painting has been shown to improve beam quality and reduce beam losses \cite{hotchi_montague_2020, saha_jpac_2025, henderson_sns_2014}, detailed control of the phase space distribution in the presence of strong space charge forces remains challenging.

In this work, we explore a novel painting method which we refer to as \textit{eigenpainting}, first proposed by Danilov et al. \cite{danilov_scbd_2003}. The method injects particles into a single mode of a coupled focusing lattice, forming a stationary distribution of zero volume in the four-dimensional (4D) transverse phase space, yet a finite area when projected onto the transverse plane. In one variant of eigenpainting, which we call \textit{uniform} eigenpainting, the painted distribution generates linear space charge forces and is an equilibrium solution to the Vlasov-Poisson equations \cite{kapchinskij_limitations_1959, chung_generalized_2016}. Both properties---zero four-dimensional emittance, linear space charge forces---have the potential to mitigate space charge effects in hadron accelerators \cite{burov_circular_2002, burov_circular_2013, cheon_spinning_2022}. More generally, eigenpainting offers a new method to control the full 4D phase space distribution of intense hadron beams.

Danilov \cite{danilov_scbd_2003} first suggested the possibility of uniform eigenpainting in the Spallation Neutron Source (SNS) accumulator ring. Holmes et al. \cite{holmes_injection_2018} performed detailed particle-in-cell (PIC) simulations to study the feasibility of the method in the presence of realistic physics effects and machine imperfections, which violate the assumptions used to derive the painting method. They found that, with several modifications to the SNS ring, the desired beam properties approximately maintained throughout accumulation: the beam developed a nonzero but small four-dimensional emittance and preserved a fairly uniform density. Motivated by these results, we report in this Letter the first experimental test of the eigenpainting method in the SNS accumulator ring.

\begin{figure*}[t]
    \centering
    \subfloat[Schematic trajectories for several painting schemes.\label{fig:painting_schemes_a}]{
    \includegraphics[width=0.34\textwidth]{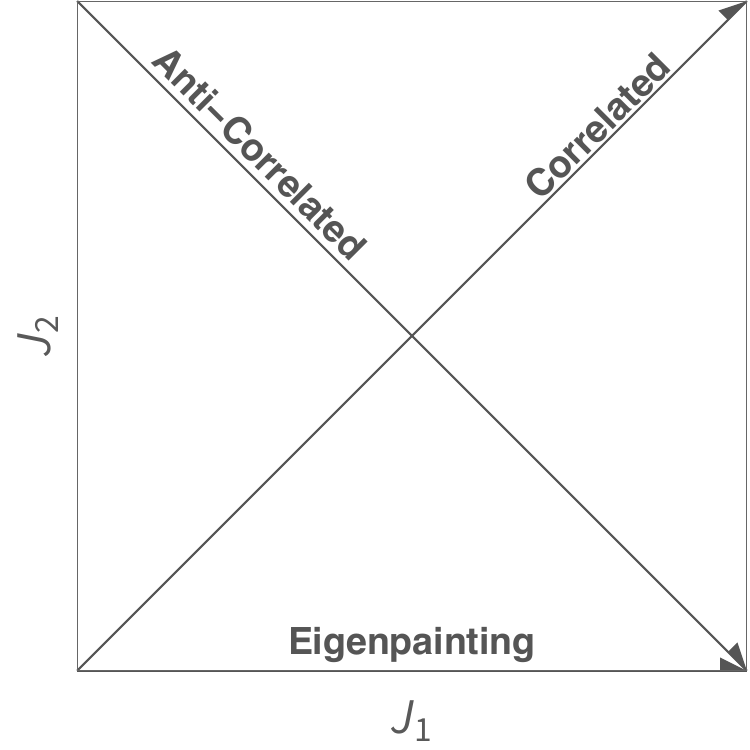}}
    ~
    \subfloat[$x$-$y$ distributions.\label{fig:painting_schemse_b}]{
    \includegraphics[width=0.60\textwidth]{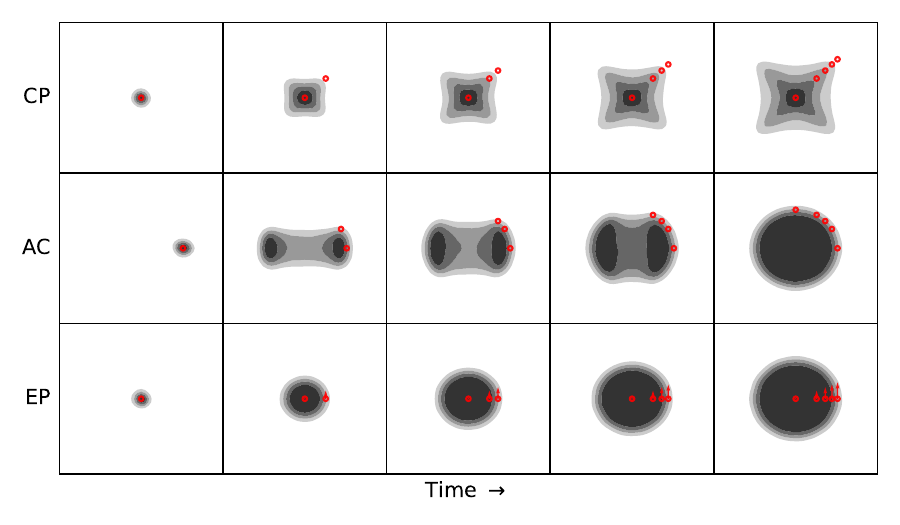}}
    \caption{Representation of the painting trajectories in the plane of mode actions, $J_{1,2}$ corresponding to \textit{correlated} (CP), \textit{anti-correlated} (AP) and \textit{eigenpainting} (EP) schemes (left) and time evolution of the $x$-$y$ distribution of beam produced by each scheme (right). Red points indicate the centroid position of injected beamlets; arrows indicate centroid angle if non-zero. }
    \label{fig:painting_schemes}
\end{figure*}

\textit{Method}---Eigenpainting is built on the eigenvector analysis of linear coupled optics \cite{edwardsteng1973,willeke1989methods,lebedev_coupled_2010}. Consider a linear, periodic focusing lattice with one-turn symplectic transfer matrix $\mathbf{M}$. We study the eigenvectors that solve $\mathbf{M} \mathbf{v} = e^{-i\mu}\mathbf{v}$. The four solutions are labeled $\mathbf{v}_1$, $\mathbf{v}_1^*$, $\mathbf{v}_2$, $\mathbf{v}_2^*$, where $*$ indicates complex conjugation; we consider only $\mathbf{v}_1$ and $\mathbf{v}_2$. When the eigenvalues are distinct, the eigenvectors define two independent modes of oscillation. We may write the phase space coordinates $\mathbf{x} = (x, x', y, y')$, where $x$ and $y$ are positions and $x'$ and $y'$ are momenta, as a linear combination of the modes:
\begin{equation}
    \mathbf{x} = \Re \left\{ \sum_{k}\sqrt{2 J_k} e^{-i \Psi_k} \mathbf{v}_k \right\},
\end{equation}
where $\Re$ selects the non-imaginary component, $\Psi_k$ is an initial phase, and $J_k$ is a single-particle invariant, reducing to the Courant-Snyder (CS) invariant \cite{courant_theory_1958} in the absence of coupling. The transfer matrix advances the phases by angle $\mu_k$:
\begin{equation}
    \mathbf{Mx} = \Re \left\{ \sum_{k}\sqrt{2 J_k} e^{-i (\Psi_k + \mu_k)} \mathbf{v}_k \right\}.
\end{equation}
A symplectic matrix $\mathbf{V}$ constructed from the eigenvectors,
\begin{equation}
    \mathbf{V} = [\Re\{\mathbf{v}_1\}, -\Im\{\mathbf{v}_1\}, \Re\{\mathbf{v}_2\}, -\Im\{\mathbf{v}_2\}],
\end{equation}
where $\Im$ selects the imaginary component, converts the transfer matrix to the form:
\begin{equation}
\begin{aligned}
    \mathbf{V}^{-1} \mathbf{M}\mathbf{V} &= \mathbf{P},
\end{aligned}
\end{equation}
where $\mathbf{P}$ is a block-diagonal matrix of $2 \times 2$ rotation matrices $\mathbf{R}(\mu_k)$ for phase advance angles $\mu_k$:
\begin{equation}
    \mathbf{P} = 
    \begin{bmatrix}
        \mathbf{R}(\mu_1) & \mathbf{0} \\ 
        \mathbf{0} & \mathbf{R}(\mu_2)
    \end{bmatrix}.
\end{equation}
In the normalized coordinates $\mathbf{u}=(u_1,u_1',u_2,u_2')=(\mathbf{u}_1,\mathbf{u_2})$ defined by $\mathbf{x} = \mathbf{V} \mathbf{u}$, the motion is uncoupled and the turn-by-turn trajectory traces a circle of area $2 \pi J_k$ in each plane, $\mathbf{u}_k$. In the unnormalized coordinates, the real component of each eigenvector traces an ellipse in each two-dimensional subspace. We therefore call the modes \textit{elliptical}, or, in special cases, \textit{circular} \cite{burov_circular_2002}. Uncoupled lattices generate \textit{planar} modes corresponding to independent $x$ and $y$ motion.

Eigenpainting follows straightforwardly from this framework. Let us first define the \textit{painting trajectory} as the time-dependent path of the injected beam centroid $\mathbf{x}_c(t)$ during accumulation.\footnote{In practice, the injection point is typically fixed and the closed orbit that defines the origin of the phase space of the ring is varied.} We may write the trajectory in terms of the eigenvectors:
\begin{equation}\label{eq:injection}
    \mathbf{x}_c(t) = 
    \Re 
    \left\{
        \sum_k
        \sqrt{2 J_k(t)} \mathbf{v}_1 e^{-i\Psi_k(t)}
    \right\},
\end{equation}
where $J_k(t)$ and $\Psi_k(t)$ are chosen time-dependent amplitudes and phases and $t$ is scaled to the range $0 \le t \le 1$. Let us select constant phases $\Psi_k(t) = \Psi_k$, so that the trajectory corresponds to a line in $J_1$-$J_2$ space. We define \textit{eigenpainting} by the choice $J_2(t) = 0$ (or $J_1(t) = 0$), which injects particles into a single mode:
\begin{equation} \label{eq:injection_eig}
{
    \mathbf{x}_c(t) = 
    \Re 
    \left\{
        \sqrt{2 J_{k}(t)} \mathbf{v}_k e^{-i\Psi_k}
    \right\}
}
\end{equation}
for chosen mode number $k$. Continuous injection at fixed amplitude generates an ellipse in the $x$-$y$ plane, with particle velocities tangent to the ellipse. Variation of the injection amplitude creates an elliptically symmetric distribution with a vortex velocity field. The distribution occupies zero volume in the four-dimensional phase space, with potential applications to luminosity boosting via round-to-flat transformations in high-energy colliders \cite{burov_circular_2013}, to beam cooling \cite{burov_circular_2002}, and to suppression of particle-core resonances \cite{cheon_spinning_2022}.

The choice $J_k(t) = \tilde{J}_k t$, where $\tilde{J}_k$ is a maximum amplitude, leads to a uniform particle density in the $x$-$y$ plane with an elliptical boundary at all times during injection; see Fig.~\ref{fig:painting_schemes}. We call this scheme \textit{uniform eigenpainting}. Remarkably, uniform eigenpainting can be adapted to generate a periodic Vlasov equilibrium distribution \cite{danilov_scbd_2003, kapchinskij_limitations_1959, lund_generation_2009, chung_generalized_2016, hoover_four_2024} in the presence of space charge forces. The painting scheme in Eq.~\eqref{eq:injection_eig} generates the following distribution function $f$ in phase space:
\begin{equation}\label{eq:danilov}
f(J_1, J_2) \propto \Theta(\tilde{J}_1 - {J}_1) \delta(J_2),
\end{equation}
where $\Theta$ is the Heaviside step function, $\delta$ is the Dirac delta function, and we take mode 1 to be the mode with non-zero amplitude. The distribution is the limit of the generalized Kapchinskij-Vladimirskij (KV) distribution \cite{kapchinskij_limitations_1959},
\begin{equation}
f(J_1, J_2) \propto \delta(J_{1} / \tilde{J}_{1} + J_{2} / \tilde{J}_{2} -1 ),
\end{equation}
as $\tilde{J}_2 \rightarrow 0$.

It is important to note that the linear defocusing force from the beam effectively modifies the one-turn transfer matrix $\mathbf{M} \rightarrow \hat{\mathbf{M}}$, where $\hat{\mathbf{M}}$ accounts for both applied and self-generated linear forces. If the beam envelope is periodic, we may compute new eigenvectors $\mathbf{v}_k \rightarrow \hat{\mathbf{v}}_k$ from $\hat{\mathbf{M}}$. Thus, to generate an equilibrium distribution with linear space charge forces, we must paint along the modified eigenvectors $\hat{\mathbf{v}}_k$, which can be computed by numerically solving a system of ordinary differential equations describing the beam envelope dynamics \cite{lund_match_2006, hoover_matched_2021}.

As a beam distribution approaches an ideal KV distribution, the tune spread approaches zero, effectively narrowing the width of resonance stopbands in the lattice and presumably unlocking larger regions of tune space and higher intensities. The reduced tune spread in this simplified incoherent picture \cite{baartman_betatron_1998, holmes_dynamics_1999} is balanced by the coherent instability of the KV distribution for certain combinations of beam perveance and applied focusing strength \cite{hoffman_stability_1983, lund_generation_2009}. Nonetheless, the dynamics of such distributions in strongly coupled focusing systems during accumulation, in different space charge regimes, has not been studied in detail and deserves careful treatment. If coherent stability prevents practical use of eigenpainting for certain operating parameters, experimental tests remain interesting to test theoretical predictions and benchmark simulations in unique scenarios.

\begin{figure*}
  \centering
  \subfloat[Calibration Data\label{subfig-1:tbt}]{
    \includegraphics[width=0.3\textwidth]{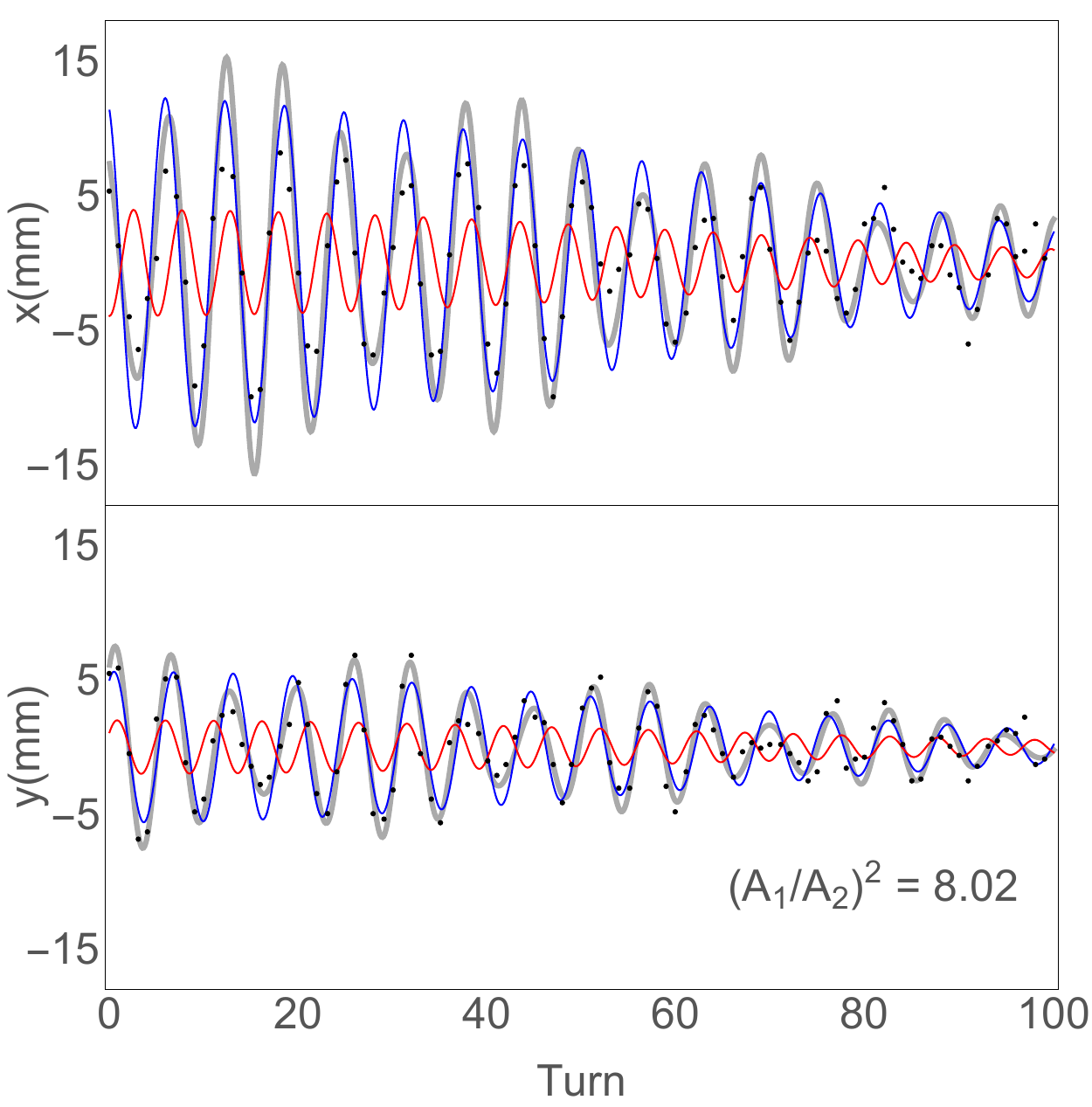}}
    ~
    \subfloat[Injection at $t=0$\label{subfig-2:tbt}]{
    \includegraphics[width=0.3\textwidth]{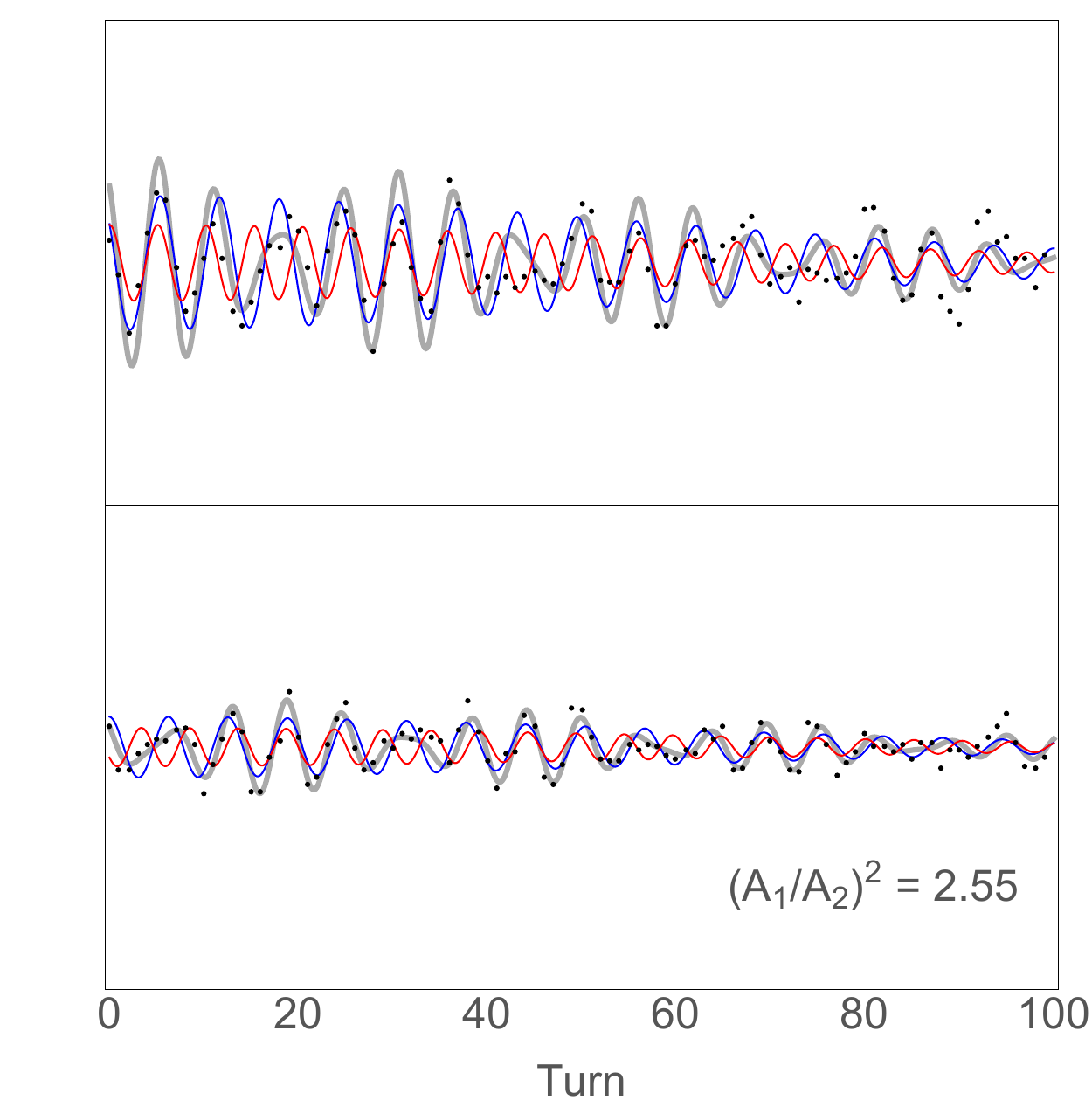}}
    \subfloat[Injection at $t=T_f$\label{subfig-3:tbt}]{
    \includegraphics[width=0.3\textwidth]{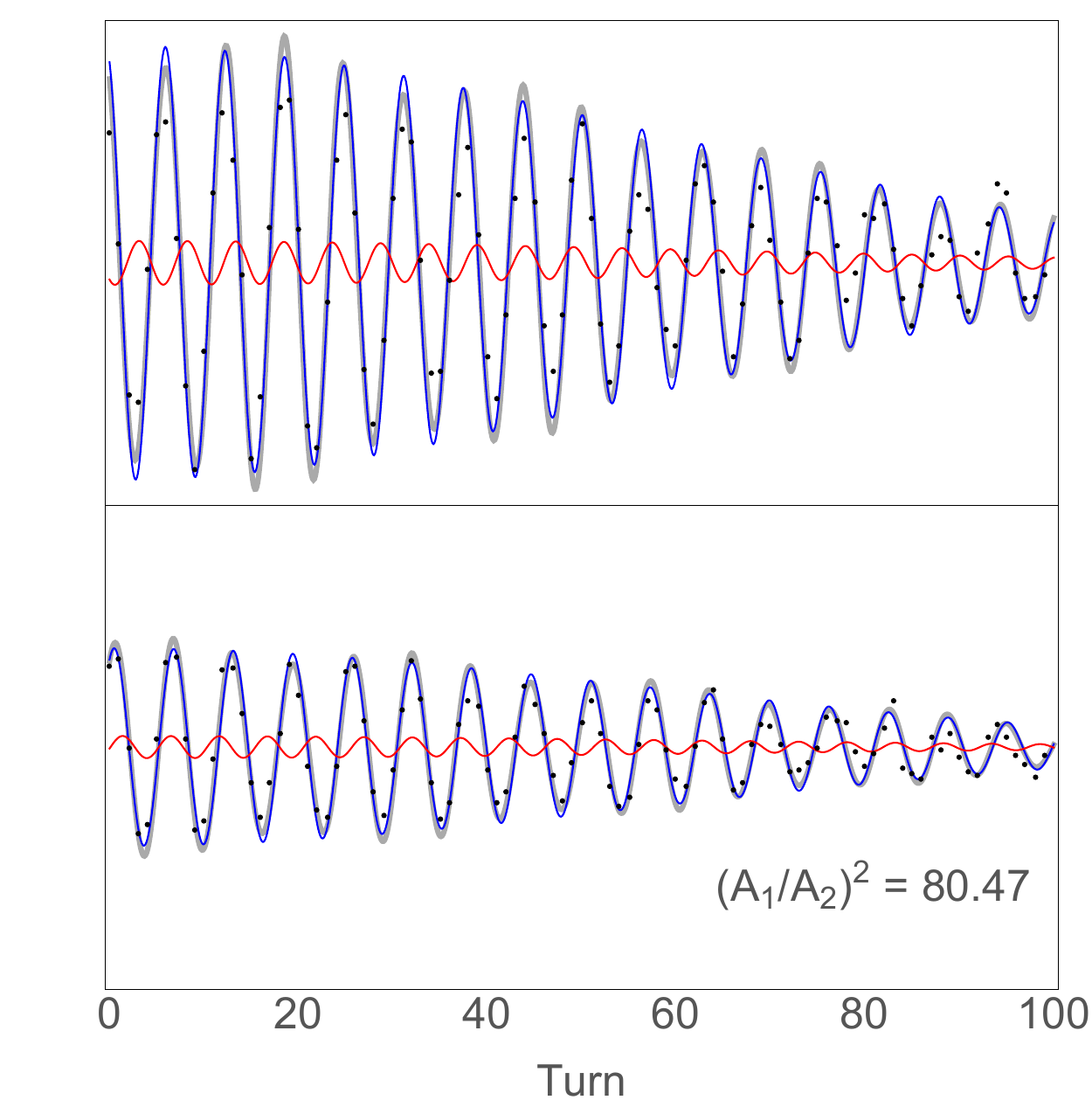}
    }~

  \caption{Turn-by-turn data (black points) for three sets of injection coordiates from the BPM in the injection straight quad doublet downstream of the foil (BPM A13), showing the fit trajectory in gray and contributions from mode 1 (blue), mode 2 (red). Offsets have been removed.}\label{fig:tbt}
\end{figure*}

\textit{Experimental Demonstration}---We performed an experiment to demonstrate uniform eigenpainting in the Spallation Neutron Source~(SNS) at Oak Ridge National Laboratory. The SNS consists of a 1.3~GeV superconducting $\text{H}^{-}$ linac, a 248~m long Accumulator Ring, and a transport line to the liquid mercury target. The ring is composed of four achromatic arcs, and four dispersion-free straight sections dedicated to injection, extraction, bunching RF, and collimation~\cite{henderson_sns_2014}. Two solenoid magnets were added to a drift in the extraction straight to facilitate non-planar modes for this research program~\cite{evans:hb2023}. The injection region contains eight time-varying dipoles to create independent horizontal and vertical four-bumps for control of the position and angle of the closed orbit at the injection location throughout the 1~ms injection cycle. To increase the phase space coverage of the injection system, these experiments were conducted at a kinetic energy of 800~MeV.  The Ring-Target Beam Transport (RTBT) line is equipped with wirescanners and sufficiently flexible optics to reconstruct the 4D distribution of the bunch extracted from the ring which allows measurement of the beam envelope, 4D emittance, and beam uniformity. 

The functional form of the painting trajectory is determined by Eq.~\eqref{eq:injection_eig}. We must find amplitude settings for the eight injection kickers that correspond to the initial and final points on the trajectory: $\mathbf{x}_c(0) = 0$ and $\mathbf{x}_c(T) = \Re \left\{ \sqrt{2 \tilde{J_1}} \mathbf{v}_{1} e^{-i\Psi_1}\right\}$. To tune the amplitude of each mode, we use the turn-by-turn centroid motion of a single beam pulse injected into the ring. Assuming a Gaussian energy spread in the presence of chromaticity, the phase space coordinates of the bunch centroid evolve according to~\cite{pelaia2016parameterestimationgaussiandampedsinusoids}:

\begin{equation}\label{eq:injectionfit}
\mathbf{x}^{j}(n) =  \sum_{k=1,2}^{} A_k e^{-\gamma_k n^2 - i (2\pi n \nu_k + \Psi_k) }\mathbf{v}_{k}^{j}(n) + \mathbf{c}^{j},
\end{equation}
where $n$ is the turn number, $k$ is the mode number, $j$ is the BPM number, $\mathbf{c}$ is the closed orbit offset, $\nu_k$ is the tune, and $\gamma_k$ is a damping coefficient. Each BPM measures the $x$ and $y$ components of Eq.~\eqref{eq:injectionfit}. The tune $\nu_k$, and damping coefficient $\gamma_k$ of each mode are the same at every BPM and depend only on the ring optics. Global parameters are estimated with Pelaia's geometrical estimation method~\cite{pelaia2016parameterestimationgaussiandampedsinusoids}, simultaneously fitting data from all BPMs. The initial amplitude $A_k=\sqrt{2 J_k}$ and phase $\Psi_k$ are fit independently for all BPM's once the global parameters have been established. The closed orbit offset $\mathbf{c}$ are distinct for each BPM and also fit independently. These values are fit using BPM data for each set of kicker amplitudes using Eq.~\eqref{eq:injectionfit} with the previously determined values of $\nu_k$, $\gamma_k$, $\mathbf{v}^j_k$. 

In the uncoupled lattice, the tunes were initially set to $\nu_x=\nu_y=0.177$. The central tunes estimated using this method indicate a closest approach of the tunes of $<10^-3$, giving an indication of the level of coupling prior to the addition of the solenoid field to the lattice. A model of the machine using the PTC~\cite{Schmidt:573082} module in MAD-X~\cite{grote2003mad,madweb} was calibrated to reproduce the observed tunes by varying quadrupole strengths. After turning on the solenoid magnets, we measured tunes $\nu_1= 0.196$ and $\nu_2=0.158$. The model solenoid strength was then varied to reproduce the observed tunes. The complex eigenvectors $\mathbf{v}^j_k$ were then extracted from the model. The $\mathbf{v}$ are normalized to the root of the local betafunctions, and include the phase from the injection point to each BPM. This allows the global amplitude $A_k$ and phase $\Psi_k$ to be easily fit independently for each injection. 

Fig.~\ref{subfig-1:tbt} shows vertical and horizontal turn-by-turn data for one $\approx10$~nC pulse of beam at one BPM used for tune calibration. Coupling is evident in the modulation of the envelope due to beating between $\nu_1$ and $\nu_2$ present in both planes. The kicker strengths were then tuned using the online OpenXAL~\cite{openxal} model to produce the injection coordinates specified by the eigenvector $\mathbf{v}_{foil}$ at the foil with an amplitude as large as the kickers would allow. We chose the mode with $x \propto +y'$ for injection, and a phase that minimizes the $x'$ to minimize losses caused by the geometry of apertures near the injection region~\cite{Holmes-Injection}.

Fig.~\ref{subfig-3:tbt} shows the turn-by-turn BPM data for the optimized single-mode injection at the end point of the painting process. Despite the coupled lattice, as $\varepsilon_2$ decreases the modulation of the envelope is reduced. The ratio of the amplitudes for this centroid analysis, $(\varepsilon_{1}/\varepsilon_{2})^2\approx80$, represents a maximum possible emittance ratio limited only by injection errors. We discuss additional sources of emittance ratio dilution later. The optimized injection in Fig.~\ref{subfig-2:tbt} corresponds to the coordinates 
$\mathbf{x}(T_f) = \left( 10.3~\text{mm}, 0.03~\text{mrad}, 2.0~\text{mm}, 0.91~\text{mrad}\right)$.

We used a similar procedure to inject onto the closed orbit at $t=0$. The deep modulation shown in Fig.~\ref{subfig-2:tbt} is due to roughly equal amplitudes in mode 1 and 2. We reduced $A_1^2$ by a factor of $\approx10$, giving $\mathbf{x}(0) = \left( 4.52~\text{mm}, -0.13~\text{mrad}, 2.37~\text{mm}, 0.014~\text{mrad}\right)$, which is non-ideal as it several beam widths from the closed orbit. During the experiment coordinates were optimized with the solenoid off due to limitations in the online model. The solenoid introduced a small deviation in the closed orbit at the foil leading to the observed amplitude. This violates one of the assumptions required to inject a Danilov distribution. However, we compared simulations of the case with no offset, and the observed value and saw virtually no difference in the final distribution, which was dominated by space charge effects. In the absence of space charge, the painting path executed leads to a hollow beam filling primarily mode 1. More robust online analysis would help reduce this error.

With the kicker settings defined at the two end points, kicker waveforms were calculated according to Eq.~\eqref{eq:injection_eig} and used to inject $8.8~\mu$C of beam over 600 turns. We used the MENT algorithm \cite{Minerbo} to reconstruct the accumulated 4D phase space distribution from 1D profile measurements of the extracted beam. We find that the selected 1D views place fairly tight constraints on the 4D distribution. A description of the measurement procedure and uncertainty quantification is found in \cite{Hoover-4d}.

\begin{figure}
    \centering
    \begin{minipage}[l]{0.05\columnwidth} 
        \centering
        \rotatebox{0}{~}
    \end{minipage}
    \begin{minipage}[l]{0.45\columnwidth} 
        \centering
        \rotatebox{0}{Mode 1}
    \end{minipage}
    \vspace{0.1in}
    \begin{minipage}[l]{0.45\columnwidth} 
        \centering
        \rotatebox{0}{Mode 2}
    \end{minipage} 
    \begin{minipage}[l]{0.05\columnwidth} 
        \centering
        \rotatebox{90}{Reconstructed}
    \end{minipage}
    \begin{minipage}[c]{0.45\columnwidth} 
        \centering
        \includegraphics[width=\columnwidth]{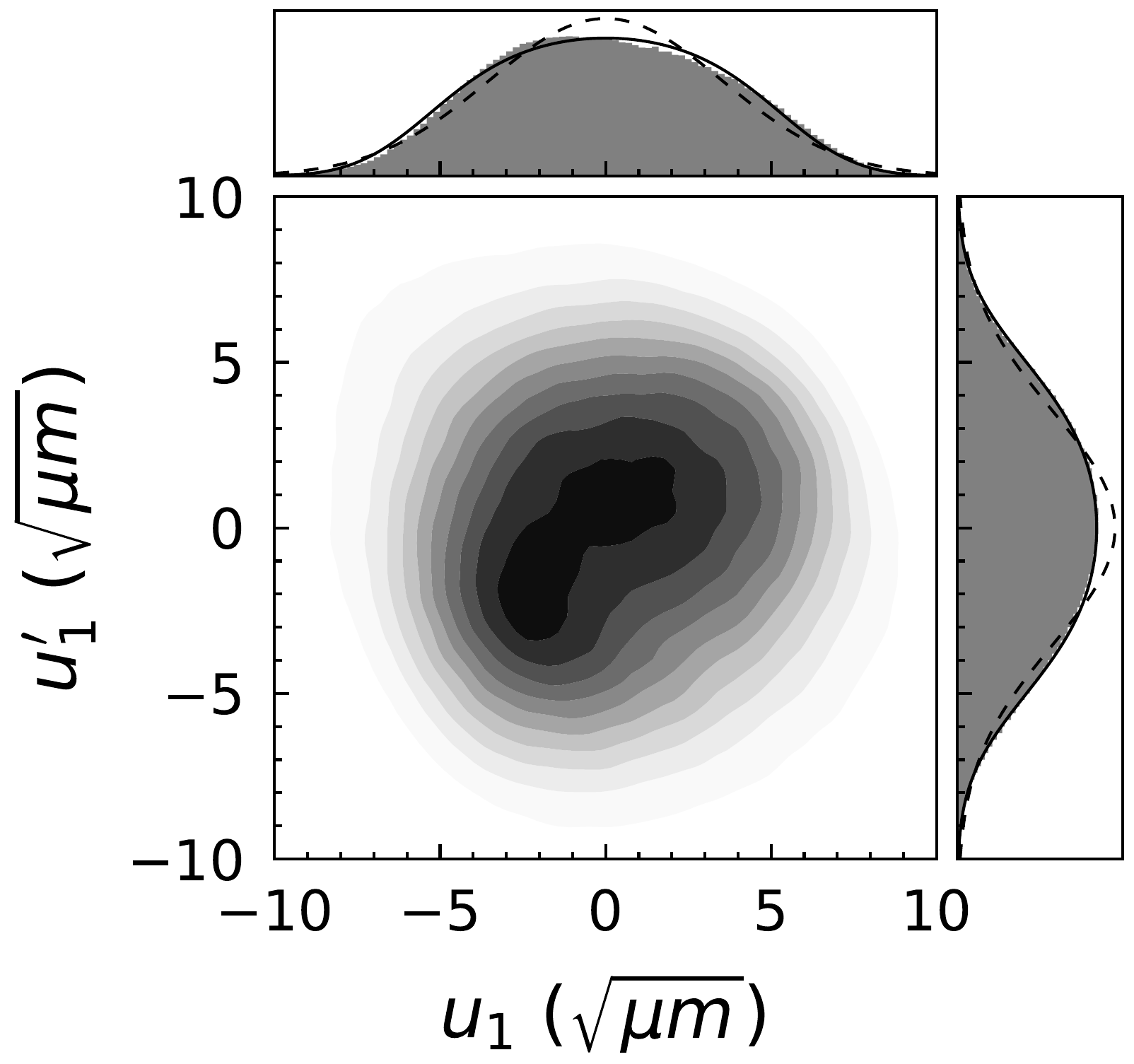} 
    \end{minipage}
    \vspace{0.1in}
    \begin{minipage}[c]{0.45\columnwidth} 
        \centering
        \includegraphics[width=\columnwidth]{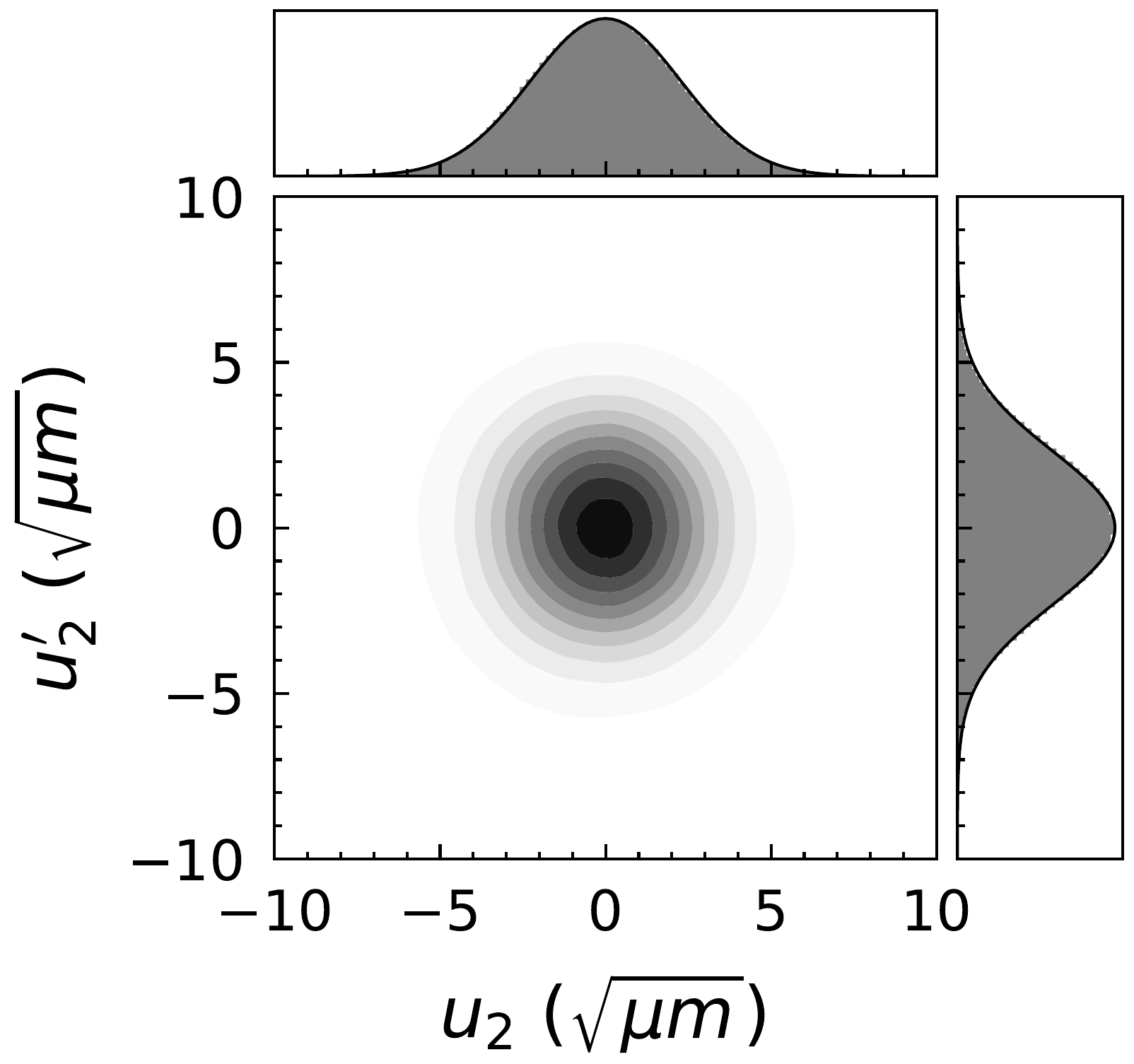} 
    \end{minipage}
    \begin{minipage}[l]{0.05\columnwidth} 
        \centering
        \rotatebox{90}{Simulated}
    \end{minipage}
    \begin{minipage}[c]{0.45\columnwidth} 
        \centering
        \includegraphics[width=\columnwidth]{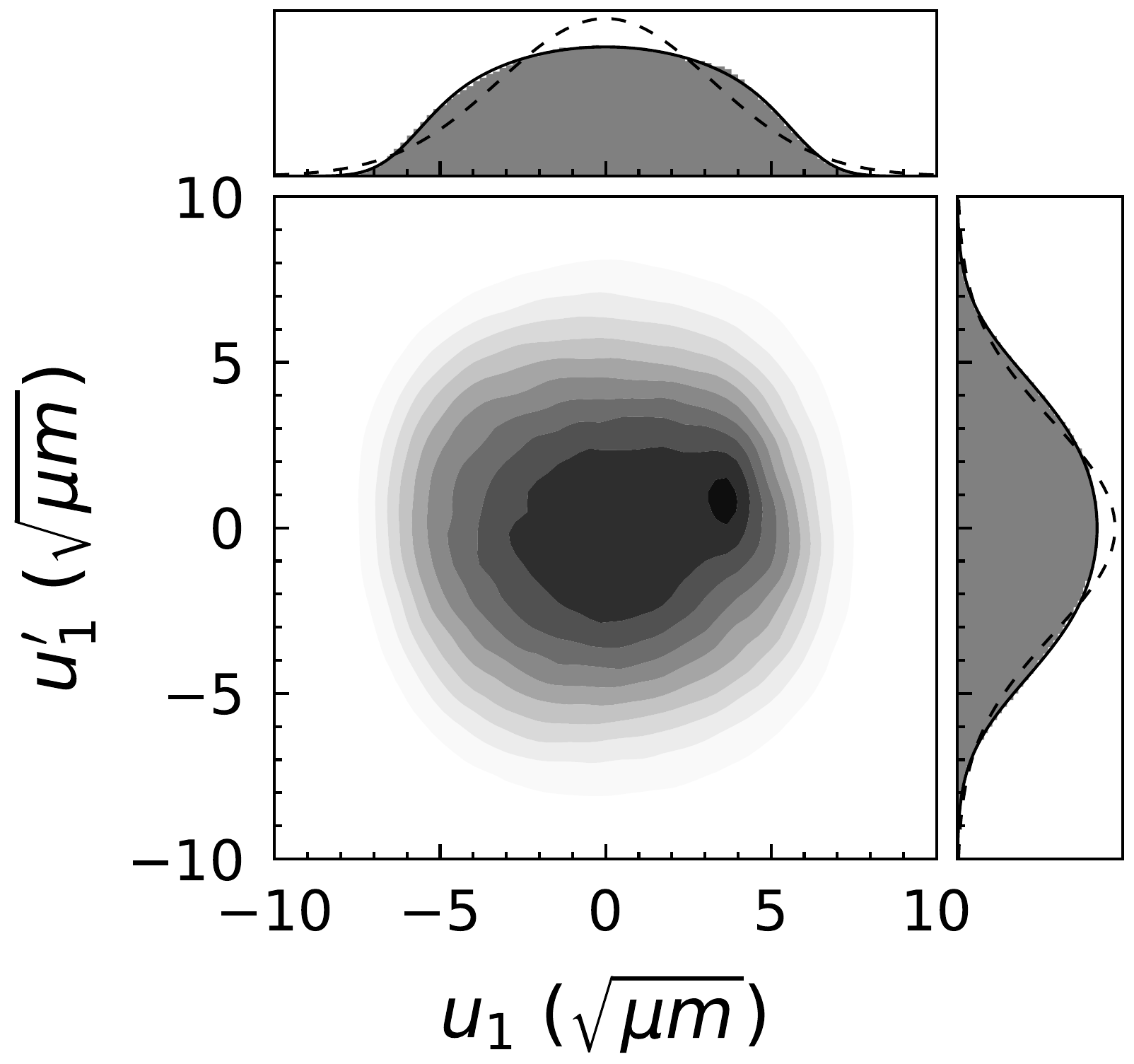} 
    \end{minipage}
    \begin{minipage}[c]{0.45\columnwidth} 
        \centering
        \includegraphics[width=\columnwidth]{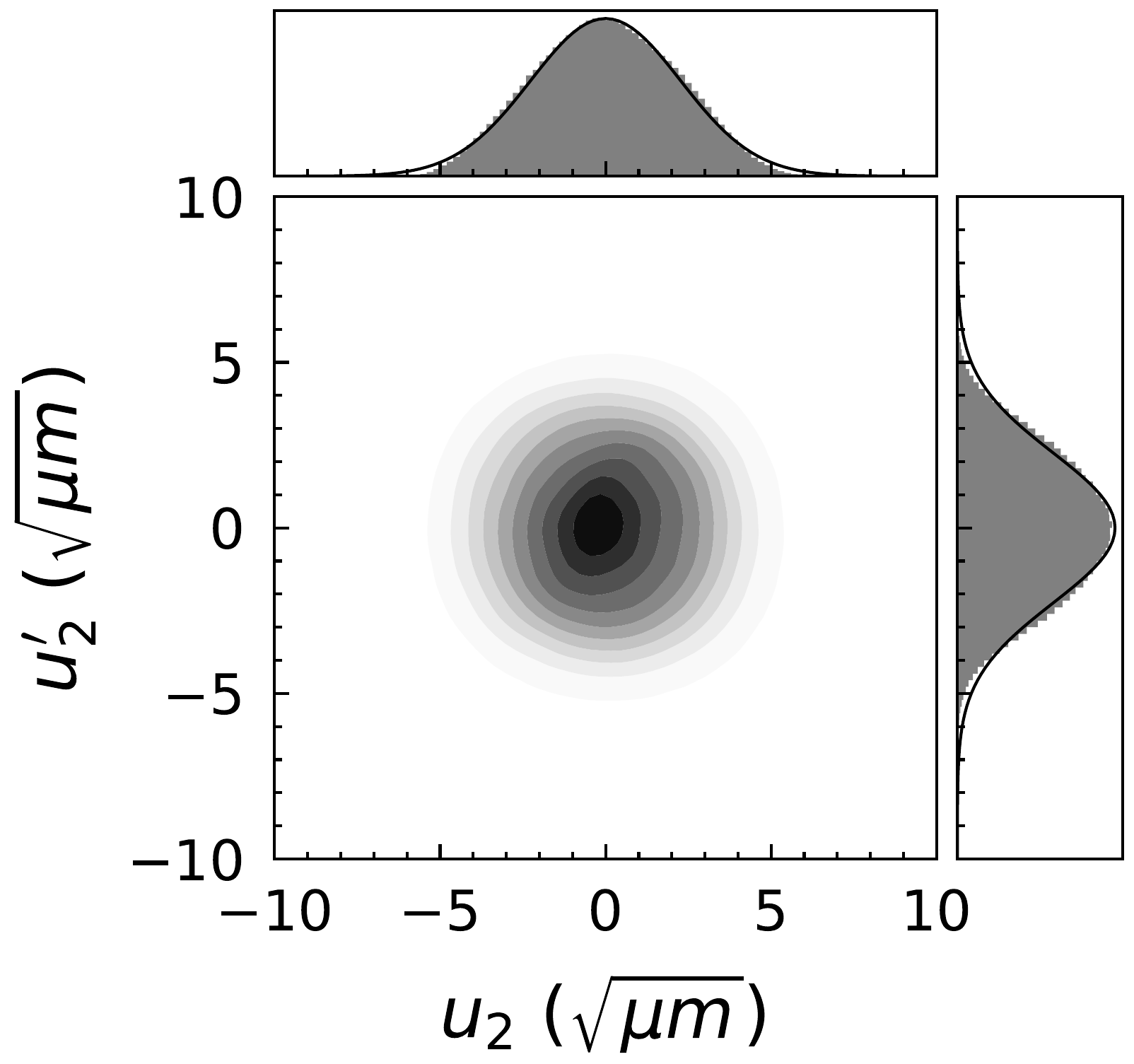} 
    \end{minipage}
    \caption{2D projections of the reconstructed (top) and simulated (bottom) phase space distributions in normalized coordinates. Solid black lines in the marginal projections represent a fit to a uniform-Gaussian convolution as described in the text. Dashed lines represent a pure Gaussian distribution.}
    \label{fig:normalized-dist}
\end{figure}

Fig.~\ref{fig:normalized-dist} shows 2D projections of the reconstructed distribution in normalized coordinates. Ideally, the Mode 1 projection would be an approximately uniform distribution within a circular boundary, while the Mode 2 projection would be a much smaller Gaussian distribution, representing the finite size of the injected minipulse. The root-mean-square (rms) area of each projection is equal to the intrinsic emittances, which we calculate to be $\varepsilon_1 = 12.4$ and $\varepsilon_2 = 5.1~\upmu$m, giving a ratio $\varepsilon_1/\varepsilon_2 = 2.4$.

Particle-in-cell (PIC) simulations of the experiment using PyORBIT~\cite{PYROBIT} are shown in Fig.~\ref{fig:normalized-dist}. 
The simulation predicted rms emittances $\varepsilon_1 = 10.3$ $\upmu$m and $\varepsilon_2 =4.5~\upmu$m, giving the ratio $\varepsilon_1/\varepsilon_2 =~2.3$, in line with the measured values. The simulated density projections are in rough agreement with the reconstructed projections: the first projection has asymmetric features and a more uniform density than the second projection, which is nearly Gaussian. The simulated distribution is independently normalized at the injection point, not the measurement point, which could account for some of the phase difference in asymmetric features.

To provide a more quantitative analysis of the measured distribution, we fit each projection to a simple model density based on the expected features. The Mode 2 projection is well-described by a 2D Gaussian distribution with standard deviation $\sigma_2 = 2.29$~$\sqrt{\upmu \text{m}}$. We model the Mode 1 projection as the convolution of a Heaviside step function of radius $R$, representing the uniformly filled mode, with a Gaussian distribution of rms width $\sigma_1$, representing the finite size of the injected beam pulses. The fit gives $\sigma_1 = 1.79$~$\sqrt{\upmu \text{m}}$ and $R = 6.03$~$\sqrt{\upmu \text{m}}$. Assuming an injected pulse emittance of $\varepsilon^{inj} = 0.25~\upmu \mathrm{m}$ in both planes, we estimate a factor of $\approx 2$ increase in the emittance during accumulation due to mismatch. While mismatch of the injected pulse accounts for some emittance growth, it cannot account for the much larger emittance growth observed in the experiment, which we attribute to nonlinear space charge forces.

A beam with the measured intrinsic emittances and rms-matched to the bare lattice optics would generate rms tune shifts of $\Delta \nu_{1,2}  = -0.28,-0.29$ in the Gaussian limit and $\Delta \nu_{1,2}  = -0.14,-0.15$ in the uniform (KV) limit. Despite the difference in emittances, the tune shift is nearly the same for the two modes, as expected for non-planar modes~\cite{burovDerbenev2009}. The simulated rms tune shifts at the end of injection are approximately $\Delta \nu_{1,2} = -(0.16, 0.15)$. For comparison, simulations without space charge indicate a chromatic tune spread of $0.006$ in both planes. Note that these tunes are calculated after significant evolution of the beam density during accumulation. With the injection amplitude used in this experiment, space charge pushes the tunes across the integer resonance early in the accumulation cycle, leading to a potentially complex dynamics and strong nonlinear emittance growth, modulated by the painting process. Changing the bare lattice tunes may be necessary to address this issue in future experiments.

Non-uniform charge density violates the linear space charge assumption used to justify self-consistent injection. Some non-uniformity is due to the finite extent of the linac beam in the $x-y$ plane. Beam in mode 2 also affects charge uniformity as the beam propagates around the ring. The finite extent of the linac beam in 4D space along with mismatch to the ring optics can both contribute motion in mode 2. Injection offsets will also contribute. Turn-by-turn data above characterizes the mode 2 motion from injection offset in the bare lattice, which is relatively small compared to the final emittance ratio. However, space charge modifies the eigenvectors which will lead to additional injection offset. We do not attempt to disentangle the contribution of offset, mismatch, and finite beam size with respect to space charge. Eliminating space charge but maintaining realistic linac beam, foil scattering, and non-linear fringe fields in simulation gives an emittance ratio of $\approx16$. This implies an emittance growth factor of $\approx6.6$ due to space charge.

\textit{Conclusion}---We have demonstrated a novel technique, which we refer to as \textit{eigenpainting}, to inject particles into a non-planar mode of a linearly coupled accelerator lattice, achieving a ratio of $\varepsilon_1/\varepsilon_2 = 2.4$. By injecting with an amplitude proportional to the square root of time, we attempted to uniformly fill the mode. Strictly, we violated several assumptions that would enable injection of a linear-force equilibrium distribution in the ring. We injected a finite sized beam near, but not on, the closed orbit, and we did not account for modified eigenvectors due to space charge throughout injection. 

We note PIC simulations reproduce the measured emittance ratio to within $\approx5\%$, as well as key features of the beam distribution in 4D phase space. These simulations also indicate that strong nonlinear space charge forces are responsible for the large 4D emittance growth observed in the experiment. To improve the painted beam quality, future work must account for the effect of space charge throughout the painting process, likely through empirical maximization of the emittance ratio. Due to the round-to-flat transformation, a circular beam with $\varepsilon_1 = 2 \epsilon_x$ has the same tune shift, or rms size, as a round, planar beam with $\varepsilon_x = \varepsilon_y$~\cite{burov_circular_2013}. At the threshold of $\varepsilon_1/\varepsilon_2 = 4$, the 4D volume of the equivalent beams is equal. For ratios larger than 4, the non-planar beam is brighter for a given tune shift. Future experiments should conservatively  target a ratio $\varepsilon_1/\varepsilon_2\gtrapprox10$ to see a practical impact.

With a sufficiently flexible injection system and modest changes to standard ring lattices, eigenpainting may provide a new technique to mitigate space charge and produce high-brightness hadron beams. Additionally, the special case of uniform eigenpainting could enable experimental benchmarking of theoretical models of space charge effects over a range of beam intensities.

\textit{Acknowledgements}---We would like to thank Charles Peters (ORNL) for help in the control room, and Sarah Cousineau (ORNL) for helpful feedback on this manuscript. We would also like to acknowledge Steve Lund (MSU) for discussions that clarified the connection between the Danilov and KV distributions.
This manuscript has been authored by UT Battelle, LLC under Contract No. DE-AC05-00OR22725 with the U.S. Department of Energy. This research used resources at the Spallation Neutron Source, a DOE Office of Science User Facility operated by the Oak Ridge National Laboratory. This work was partially funded by Field Work Proposal ORNL-ERKCS41-Funding.

\textit{Data availability}---The data that support the findings of this article are openly available~\cite{dataset}.

\appendix

\bibliographystyle{apsrev4-1}
% \bibliography{main}

\begin{thebibliography}{41}%
\makeatletter
\providecommand \@ifxundefined [1]{%
 \@ifx{#1\undefined}
}%
\providecommand \@ifnum [1]{%
 \ifnum #1\expandafter \@firstoftwo
 \else \expandafter \@secondoftwo
 \fi
}%
\providecommand \@ifx [1]{%
 \ifx #1\expandafter \@firstoftwo
 \else \expandafter \@secondoftwo
 \fi
}%
\providecommand \natexlab [1]{#1}%
\providecommand \enquote  [1]{``#1''}%
\providecommand \bibnamefont  [1]{#1}%
\providecommand \bibfnamefont [1]{#1}%
\providecommand \citenamefont [1]{#1}%
\providecommand \href@noop [0]{\@secondoftwo}%
\providecommand \href [0]{\begingroup \@sanitize@url \@href}%
\providecommand \@href[1]{\@@startlink{#1}\@@href}%
\providecommand \@@href[1]{\endgroup#1\@@endlink}%
\providecommand \@sanitize@url [0]{\catcode `\\12\catcode `\$12\catcode
  `\&12\catcode `\#12\catcode `\^12\catcode `\_12\catcode `\%12\relax}%
\providecommand \@@startlink[1]{}%
\providecommand \@@endlink[0]{}%
\providecommand \url  [0]{\begingroup\@sanitize@url \@url }%
\providecommand \@url [1]{\endgroup\@href {#1}{\urlprefix }}%
\providecommand \urlprefix  [0]{URL }%
\providecommand \Eprint [0]{\href }%
\providecommand \doibase [0]{http://dx.doi.org/}%
\providecommand \selectlanguage [0]{\@gobble}%
\providecommand \bibinfo  [0]{\@secondoftwo}%
\providecommand \bibfield  [0]{\@secondoftwo}%
\providecommand \translation [1]{[#1]}%
\providecommand \BibitemOpen [0]{}%
\providecommand \bibitemStop [0]{}%
\providecommand \bibitemNoStop [0]{.\EOS\space}%
\providecommand \EOS [0]{\spacefactor3000\relax}%
\providecommand \BibitemShut  [1]{\csname bibitem#1\endcsname}%
\let\auto@bib@innerbib\@empty
%</preamble>
\bibitem [{\citenamefont {Cousineau}(2015)}]{cousineau_high_2015}%
  \BibitemOpen
  \bibfield  {author} {\bibinfo {author} {\bibfnamefont {S.}~\bibnamefont
  {Cousineau}},\ }\href@noop {} {\bibfield  {journal} {\bibinfo  {journal}
  {Proc. of IPAC’15}\ ,\ \bibinfo {pages} {4102}} (\bibinfo {year}
  {2015})}\BibitemShut {NoStop}%
\bibitem [{\citenamefont {Franchetti}\ \emph {et~al.}(2010)\citenamefont
  {Franchetti}, \citenamefont {Chorniy}, \citenamefont {Hofmann}, \citenamefont
  {Bayer}, \citenamefont {Becker}, \citenamefont {Forck}, \citenamefont
  {Giacomini}, \citenamefont {Kirk}, \citenamefont {Mohite}, \citenamefont
  {Omet}, \citenamefont {Parfenova},\ and\ \citenamefont
  {Sch\"utt}}]{franchetti_resonance_2010}%
  \BibitemOpen
  \bibfield  {author} {\bibinfo {author} {\bibfnamefont {G.}~\bibnamefont
  {Franchetti}}, \bibinfo {author} {\bibfnamefont {O.}~\bibnamefont {Chorniy}},
  \bibinfo {author} {\bibfnamefont {I.}~\bibnamefont {Hofmann}}, \bibinfo
  {author} {\bibfnamefont {W.}~\bibnamefont {Bayer}}, \bibinfo {author}
  {\bibfnamefont {F.}~\bibnamefont {Becker}}, \bibinfo {author} {\bibfnamefont
  {P.}~\bibnamefont {Forck}}, \bibinfo {author} {\bibfnamefont
  {T.}~\bibnamefont {Giacomini}}, \bibinfo {author} {\bibfnamefont
  {M.}~\bibnamefont {Kirk}}, \bibinfo {author} {\bibfnamefont {T.}~\bibnamefont
  {Mohite}}, \bibinfo {author} {\bibfnamefont {C.}~\bibnamefont {Omet}},
  \bibinfo {author} {\bibfnamefont {A.}~\bibnamefont {Parfenova}}, \ and\
  \bibinfo {author} {\bibfnamefont {P.}~\bibnamefont {Sch\"utt}},\ }\href
  {\doibase 10.1103/PhysRevSTAB.13.114203} {\bibfield  {journal} {\bibinfo
  {journal} {Phys. Rev. ST Accel. Beams}\ }\textbf {\bibinfo {volume} {13}},\
  \bibinfo {pages} {114203} (\bibinfo {year} {2010})}\BibitemShut {NoStop}%
\bibitem [{\citenamefont {Holmes}\ \emph {et~al.}(1999)\citenamefont {Holmes},
  \citenamefont {Danilov}, \citenamefont {Galambos}, \citenamefont {Jeon},\
  and\ \citenamefont {Olsen}}]{holmes_dynamics_1999}%
  \BibitemOpen
  \bibfield  {author} {\bibinfo {author} {\bibfnamefont {J.~A.}\ \bibnamefont
  {Holmes}}, \bibinfo {author} {\bibfnamefont {V.~V.}\ \bibnamefont {Danilov}},
  \bibinfo {author} {\bibfnamefont {J.~D.}\ \bibnamefont {Galambos}}, \bibinfo
  {author} {\bibfnamefont {D.}~\bibnamefont {Jeon}}, \ and\ \bibinfo {author}
  {\bibfnamefont {D.~K.}\ \bibnamefont {Olsen}},\ }\href {\doibase
  10.1103/PhysRevSTAB.2.114202} {\bibfield  {journal} {\bibinfo  {journal}
  {Phys. Rev. ST Accel. Beams}\ }\textbf {\bibinfo {volume} {2}},\ \bibinfo
  {pages} {114202} (\bibinfo {year} {1999})}\BibitemShut {NoStop}%
\bibitem [{\citenamefont {Liu}\ \emph {et~al.}(2025)\citenamefont {Liu},
  \citenamefont {Yao}, \citenamefont {Zheng},\ and\ \citenamefont
  {Yuan}}]{liu_structure_2024}%
  \BibitemOpen
  \bibfield  {author} {\bibinfo {author} {\bibfnamefont {X.-Y.}\ \bibnamefont
  {Liu}}, \bibinfo {author} {\bibfnamefont {H.-J.}\ \bibnamefont {Yao}},
  \bibinfo {author} {\bibfnamefont {S.-X.}\ \bibnamefont {Zheng}}, \ and\
  \bibinfo {author} {\bibfnamefont {Y.-S.}\ \bibnamefont {Yuan}},\ }\href
  {\doibase 10.1103/PhysRevAccelBeams.28.054201} {\bibfield  {journal}
  {\bibinfo  {journal} {Phys. Rev. Accel. Beams}\ }\textbf {\bibinfo {volume}
  {28}},\ \bibinfo {pages} {054201} (\bibinfo {year} {2025})}\BibitemShut
  {NoStop}%
\bibitem [{\citenamefont {Hofmann}(1998)}]{hofmann_anisotropic_1998}%
  \BibitemOpen
  \bibfield  {author} {\bibinfo {author} {\bibfnamefont {I.}~\bibnamefont
  {Hofmann}},\ }\href {\doibase 10.1103/PhysRevE.57.4713} {\bibfield  {journal}
  {\bibinfo  {journal} {Phys. Rev. E}\ }\textbf {\bibinfo {volume} {57}},\
  \bibinfo {pages} {4713} (\bibinfo {year} {1998})}\BibitemShut {NoStop}%
\bibitem [{\citenamefont {Hofmann}\ \emph {et~al.}(2021)\citenamefont
  {Hofmann}, \citenamefont {Oeftiger},\ and\ \citenamefont
  {Boine-Frankenheim}}]{oeftiger_dynamics_2021}%
  \BibitemOpen
  \bibfield  {author} {\bibinfo {author} {\bibfnamefont {I.}~\bibnamefont
  {Hofmann}}, \bibinfo {author} {\bibfnamefont {A.}~\bibnamefont {Oeftiger}}, \
  and\ \bibinfo {author} {\bibfnamefont {O.}~\bibnamefont
  {Boine-Frankenheim}},\ }\href {\doibase 10.1103/PhysRevAccelBeams.24.024201}
  {\bibfield  {journal} {\bibinfo  {journal} {Phys. Rev. Accel. Beams}\
  }\textbf {\bibinfo {volume} {24}},\ \bibinfo {pages} {024201} (\bibinfo
  {year} {2021})}\BibitemShut {NoStop}%
\bibitem [{\citenamefont {Burov}(2013)}]{burov_circular_2013}%
  \BibitemOpen
  \bibfield  {author} {\bibinfo {author} {\bibfnamefont {A.}~\bibnamefont
  {Burov}},\ }\href {\doibase 10.1103/PhysRevSTAB.16.061002} {\bibfield
  {journal} {\bibinfo  {journal} {Phys. Rev. ST Accel. Beams}\ }\textbf
  {\bibinfo {volume} {16}},\ \bibinfo {pages} {061002} (\bibinfo {year}
  {2013})}\BibitemShut {NoStop}%
\bibitem [{\citenamefont {Oeftiger}\ and\ \citenamefont
  {Boine-Frankenheim}(2024)}]{oeftiger_pulsed_2024}%
  \BibitemOpen
  \bibfield  {author} {\bibinfo {author} {\bibfnamefont {A.}~\bibnamefont
  {Oeftiger}}\ and\ \bibinfo {author} {\bibfnamefont {O.}~\bibnamefont
  {Boine-Frankenheim}},\ }\href {\doibase 10.1103/PhysRevLett.132.175001}
  {\bibfield  {journal} {\bibinfo  {journal} {Phys. Rev. Lett.}\ }\textbf
  {\bibinfo {volume} {132}},\ \bibinfo {pages} {175001} (\bibinfo {year}
  {2024})}\BibitemShut {NoStop}%
\bibitem [{\citenamefont {Danilov}\ and\ \citenamefont
  {Nagaitsev}(2010)}]{danilov_nio_2010}%
  \BibitemOpen
  \bibfield  {author} {\bibinfo {author} {\bibfnamefont {V.}~\bibnamefont
  {Danilov}}\ and\ \bibinfo {author} {\bibfnamefont {S.}~\bibnamefont
  {Nagaitsev}},\ }\href {\doibase 10.1103/PhysRevSTAB.13.084002} {\bibfield
  {journal} {\bibinfo  {journal} {Phys. Rev. ST Accel. Beams}\ }\textbf
  {\bibinfo {volume} {13}},\ \bibinfo {pages} {084002} (\bibinfo {year}
  {2010})}\BibitemShut {NoStop}%
\bibitem [{\citenamefont {Henderson}\ \emph {et~al.}(2014)\citenamefont
  {Henderson} \emph {et~al.}}]{henderson_sns_2014}%
  \BibitemOpen
  \bibfield  {author} {\bibinfo {author} {\bibfnamefont {S.}~\bibnamefont
  {Henderson}} \emph {et~al.},\ }\href {\doibase
  https://doi.org/10.1016/j.nima.2014.03.067} {\bibfield  {journal} {\bibinfo
  {journal} {Nuclear Instruments and Methods in Physics Research Section A:
  Accelerators, Spectrometers, Detectors and Associated Equipment}\ }\textbf
  {\bibinfo {volume} {763}},\ \bibinfo {pages} {610} (\bibinfo {year}
  {2014})}\BibitemShut {NoStop}%
\bibitem [{\citenamefont {Yoshimoto}\ \emph {et~al.}(2006)\citenamefont
  {Yoshimoto}, \citenamefont {Irie}, \citenamefont {Kamiya}, \citenamefont
  {Kinsho}, \citenamefont {Noda}, \citenamefont {Saha}, \citenamefont
  {Takayanagi}, \citenamefont {Takeda},\ and\ \citenamefont
  {Watanabe}}]{Yoshimoto:2006wv}%
  \BibitemOpen
  \bibfield  {author} {\bibinfo {author} {\bibfnamefont {M.}~\bibnamefont
  {Yoshimoto}}, \bibinfo {author} {\bibfnamefont {Y.}~\bibnamefont {Irie}},
  \bibinfo {author} {\bibfnamefont {J.}~\bibnamefont {Kamiya}}, \bibinfo
  {author} {\bibfnamefont {M.}~\bibnamefont {Kinsho}}, \bibinfo {author}
  {\bibfnamefont {F.}~\bibnamefont {Noda}}, \bibinfo {author} {\bibfnamefont
  {P.}~\bibnamefont {Saha}}, \bibinfo {author} {\bibfnamefont {T.}~\bibnamefont
  {Takayanagi}}, \bibinfo {author} {\bibfnamefont {O.}~\bibnamefont {Takeda}},
  \ and\ \bibinfo {author} {\bibfnamefont {M.}~\bibnamefont {Watanabe}},\
  }\href@noop {} {\bibfield  {journal} {\bibinfo  {journal} {Conf. Proc. C}\
  }\textbf {\bibinfo {volume} {060626}},\ \bibinfo {pages} {1765} (\bibinfo
  {year} {2006})}\BibitemShut {NoStop}%
\bibitem [{\citenamefont {Hotchi}(2020)}]{hotchi_montague_2020}%
  \BibitemOpen
  \bibfield  {author} {\bibinfo {author} {\bibfnamefont {H.}~\bibnamefont
  {Hotchi}},\ }\href {\doibase 10.1103/PhysRevAccelBeams.23.050401} {\bibfield
  {journal} {\bibinfo  {journal} {Phys. Rev. Accel. Beams}\ }\textbf {\bibinfo
  {volume} {23}},\ \bibinfo {pages} {050401} (\bibinfo {year}
  {2020})}\BibitemShut {NoStop}%
\bibitem [{\citenamefont {Saha}\ \emph {et~al.}(2025)\citenamefont {Saha},
  \citenamefont {Harada}, \citenamefont {Tamura}, \citenamefont {Okabe},
  \citenamefont {Yoshimoto}, \citenamefont {Shobuda}, \citenamefont {Okita},
  \citenamefont {Kojima}, \citenamefont {Nakanoya}, \citenamefont {Hatakeyama},
  \citenamefont {Takayanagi}, \citenamefont {Yamamoto},\ and\ \citenamefont
  {Hotchi}}]{saha_jpac_2025}%
  \BibitemOpen
  \bibfield  {author} {\bibinfo {author} {\bibfnamefont {P.~K.}\ \bibnamefont
  {Saha}}, \bibinfo {author} {\bibfnamefont {H.}~\bibnamefont {Harada}},
  \bibinfo {author} {\bibfnamefont {F.}~\bibnamefont {Tamura}}, \bibinfo
  {author} {\bibfnamefont {K.}~\bibnamefont {Okabe}}, \bibinfo {author}
  {\bibfnamefont {M.}~\bibnamefont {Yoshimoto}}, \bibinfo {author}
  {\bibfnamefont {Y.}~\bibnamefont {Shobuda}}, \bibinfo {author} {\bibfnamefont
  {H.}~\bibnamefont {Okita}}, \bibinfo {author} {\bibfnamefont
  {K.}~\bibnamefont {Kojima}}, \bibinfo {author} {\bibfnamefont
  {T.}~\bibnamefont {Nakanoya}}, \bibinfo {author} {\bibfnamefont
  {S.}~\bibnamefont {Hatakeyama}}, \bibinfo {author} {\bibfnamefont
  {T.}~\bibnamefont {Takayanagi}}, \bibinfo {author} {\bibfnamefont
  {K.}~\bibnamefont {Yamamoto}}, \ and\ \bibinfo {author} {\bibfnamefont
  {H.}~\bibnamefont {Hotchi}},\ }\href {\doibase 10.1103/tbyh-jcq3} {\bibfield
  {journal} {\bibinfo  {journal} {Phys. Rev. Accel. Beams}\ }\textbf {\bibinfo
  {volume} {28}},\ \bibinfo {pages} {074201} (\bibinfo {year}
  {2025})}\BibitemShut {NoStop}%
\bibitem [{\citenamefont {Danilov}\ \emph {et~al.}(2003)\citenamefont
  {Danilov}, \citenamefont {Cousineau}, \citenamefont {Henderson},\ and\
  \citenamefont {Holmes}}]{danilov_scbd_2003}%
  \BibitemOpen
  \bibfield  {author} {\bibinfo {author} {\bibfnamefont {V.}~\bibnamefont
  {Danilov}}, \bibinfo {author} {\bibfnamefont {S.}~\bibnamefont {Cousineau}},
  \bibinfo {author} {\bibfnamefont {S.}~\bibnamefont {Henderson}}, \ and\
  \bibinfo {author} {\bibfnamefont {J.}~\bibnamefont {Holmes}},\ }\href
  {\doibase 10.1103/PhysRevSTAB.6.094202} {\bibfield  {journal} {\bibinfo
  {journal} {Phys. Rev. ST Accel. Beams}\ }\textbf {\bibinfo {volume} {6}},\
  \bibinfo {pages} {094202} (\bibinfo {year} {2003})}\BibitemShut {NoStop}%
\bibitem [{\citenamefont {Kapchinskij}\ and\ \citenamefont
  {Vladimirskij}(1959)}]{kapchinskij_limitations_1959}%
  \BibitemOpen
  \bibfield  {author} {\bibinfo {author} {\bibfnamefont {I.~M.}\ \bibnamefont
  {Kapchinskij}}\ and\ \bibinfo {author} {\bibfnamefont {V.~V.}\ \bibnamefont
  {Vladimirskij}},\ }in\ \href@noop {} {\emph {\bibinfo {booktitle} {{2nd
  International Conference on High-Energy Accelerators}}}}\ (\bibinfo {year}
  {1959})\ pp.\ \bibinfo {pages} {274--287}\BibitemShut {NoStop}%
\bibitem [{\citenamefont {Chung}\ \emph {et~al.}(2016)\citenamefont {Chung},
  \citenamefont {Qin}, \citenamefont {Davidson}, \citenamefont {Groening},\
  and\ \citenamefont {Xiao}}]{chung_generalized_2016}%
  \BibitemOpen
  \bibfield  {author} {\bibinfo {author} {\bibfnamefont {M.}~\bibnamefont
  {Chung}}, \bibinfo {author} {\bibfnamefont {H.}~\bibnamefont {Qin}}, \bibinfo
  {author} {\bibfnamefont {R.~C.}\ \bibnamefont {Davidson}}, \bibinfo {author}
  {\bibfnamefont {L.}~\bibnamefont {Groening}}, \ and\ \bibinfo {author}
  {\bibfnamefont {C.}~\bibnamefont {Xiao}},\ }\href {\doibase
  10.1103/PhysRevLett.117.224801} {\bibfield  {journal} {\bibinfo  {journal}
  {Phys. Rev. Lett.}\ }\textbf {\bibinfo {volume} {117}},\ \bibinfo {pages}
  {224801} (\bibinfo {year} {2016})}\BibitemShut {NoStop}%
\bibitem [{\citenamefont {Burov}\ \emph {et~al.}(2002)\citenamefont {Burov},
  \citenamefont {Nagaitsev},\ and\ \citenamefont
  {Derbenev}}]{burov_circular_2002}%
  \BibitemOpen
  \bibfield  {author} {\bibinfo {author} {\bibfnamefont {A.}~\bibnamefont
  {Burov}}, \bibinfo {author} {\bibfnamefont {S.}~\bibnamefont {Nagaitsev}}, \
  and\ \bibinfo {author} {\bibfnamefont {Y.}~\bibnamefont {Derbenev}},\ }\href
  {\doibase 10.1103/PhysRevE.66.016503} {\bibfield  {journal} {\bibinfo
  {journal} {Phys. Rev. E}\ }\textbf {\bibinfo {volume} {66}},\ \bibinfo
  {pages} {016503} (\bibinfo {year} {2002})}\BibitemShut {NoStop}%
\bibitem [{\citenamefont {Cheon}\ \emph {et~al.}(2022)\citenamefont {Cheon},
  \citenamefont {Moon}, \citenamefont {Chung},\ and\ \citenamefont
  {Jeon}}]{cheon_spinning_2022}%
  \BibitemOpen
  \bibfield  {author} {\bibinfo {author} {\bibfnamefont {Y.-L.}\ \bibnamefont
  {Cheon}}, \bibinfo {author} {\bibfnamefont {S.-H.}\ \bibnamefont {Moon}},
  \bibinfo {author} {\bibfnamefont {M.}~\bibnamefont {Chung}}, \ and\ \bibinfo
  {author} {\bibfnamefont {D.-O.}\ \bibnamefont {Jeon}},\ }\href {\doibase
  10.1103/PhysRevAccelBeams.25.064002} {\bibfield  {journal} {\bibinfo
  {journal} {Phys. Rev. Accel. Beams}\ }\textbf {\bibinfo {volume} {25}},\
  \bibinfo {pages} {064002} (\bibinfo {year} {2022})}\BibitemShut {NoStop}%
\bibitem [{\citenamefont {Holmes}\ \emph
  {et~al.}(2018{\natexlab{a}})\citenamefont {Holmes}, \citenamefont {Gorlov},
  \citenamefont {Evans}, \citenamefont {Plum},\ and\ \citenamefont
  {Cousineau}}]{holmes_injection_2018}%
  \BibitemOpen
  \bibfield  {author} {\bibinfo {author} {\bibfnamefont {J.~A.}\ \bibnamefont
  {Holmes}}, \bibinfo {author} {\bibfnamefont {T.}~\bibnamefont {Gorlov}},
  \bibinfo {author} {\bibfnamefont {N.~J.}\ \bibnamefont {Evans}}, \bibinfo
  {author} {\bibfnamefont {M.}~\bibnamefont {Plum}}, \ and\ \bibinfo {author}
  {\bibfnamefont {S.}~\bibnamefont {Cousineau}},\ }\href {\doibase
  10.1103/PhysRevAccelBeams.21.124403} {\bibfield  {journal} {\bibinfo
  {journal} {Phys. Rev. Accel. Beams}\ }\textbf {\bibinfo {volume} {21}},\
  \bibinfo {pages} {124403} (\bibinfo {year} {2018}{\natexlab{a}})}\BibitemShut
  {NoStop}%
\bibitem [{\citenamefont {Edwards}\ and\ \citenamefont
  {Teng}(1973)}]{edwardsteng1973}%
  \BibitemOpen
  \bibfield  {author} {\bibinfo {author} {\bibfnamefont {D.~A.}\ \bibnamefont
  {Edwards}}\ and\ \bibinfo {author} {\bibfnamefont {L.~C.}\ \bibnamefont
  {Teng}},\ }\href {\doibase 10.1109/TNS.1973.4327279} {\bibfield  {journal}
  {\bibinfo  {journal} {IEEE Transactions on Nuclear Science}\ }\textbf
  {\bibinfo {volume} {20}},\ \bibinfo {pages} {885} (\bibinfo {year}
  {1973})}\BibitemShut {NoStop}%
\bibitem [{\citenamefont {Willeke}\ and\ \citenamefont
  {Ripken}(1989)}]{willeke1989methods}%
  \BibitemOpen
  \bibfield  {author} {\bibinfo {author} {\bibfnamefont {F.}~\bibnamefont
  {Willeke}}\ and\ \bibinfo {author} {\bibfnamefont {G.}~\bibnamefont
  {Ripken}},\ }in\ \href@noop {} {\emph {\bibinfo {booktitle} {AIP Conference
  Proceedings}}},\ Vol.\ \bibinfo {volume} {184}\ (\bibinfo {organization}
  {American Institute of Physics},\ \bibinfo {year} {1989})\ pp.\ \bibinfo
  {pages} {758--819}\BibitemShut {NoStop}%
\bibitem [{\citenamefont {Lebedev}\ and\ \citenamefont
  {Bogacz}(2010)}]{lebedev_coupled_2010}%
  \BibitemOpen
  \bibfield  {author} {\bibinfo {author} {\bibfnamefont {V.~A.}\ \bibnamefont
  {Lebedev}}\ and\ \bibinfo {author} {\bibfnamefont {S.~A.}\ \bibnamefont
  {Bogacz}},\ }\href {\doibase 10.1088/1748-0221/5/10/P10010} {\bibfield
  {journal} {\bibinfo  {journal} {Journal of Instrumentation}\ }\textbf
  {\bibinfo {volume} {5}},\ \bibinfo {pages} {P10010} (\bibinfo {year}
  {2010})}\BibitemShut {NoStop}%
\bibitem [{\citenamefont {Courant}\ and\ \citenamefont
  {Snyder}(1958)}]{courant_theory_1958}%
  \BibitemOpen
  \bibfield  {author} {\bibinfo {author} {\bibfnamefont {E.~D.}\ \bibnamefont
  {Courant}}\ and\ \bibinfo {author} {\bibfnamefont {H.~S.}\ \bibnamefont
  {Snyder}},\ }\href@noop {} {\bibfield  {journal} {\bibinfo  {journal} {Annals
  of physics}\ }\textbf {\bibinfo {volume} {3}},\ \bibinfo {pages} {1}
  (\bibinfo {year} {1958})}\BibitemShut {NoStop}%
\bibitem [{\citenamefont {Lund}\ \emph {et~al.}(2009)\citenamefont {Lund},
  \citenamefont {Kikuchi},\ and\ \citenamefont
  {Davidson}}]{lund_generation_2009}%
  \BibitemOpen
  \bibfield  {author} {\bibinfo {author} {\bibfnamefont {S.~M.}\ \bibnamefont
  {Lund}}, \bibinfo {author} {\bibfnamefont {T.}~\bibnamefont {Kikuchi}}, \
  and\ \bibinfo {author} {\bibfnamefont {R.~C.}\ \bibnamefont {Davidson}},\
  }\href {\doibase 10.1103/PhysRevSTAB.12.114801} {\bibfield  {journal}
  {\bibinfo  {journal} {Phys. Rev. ST Accel. Beams}\ }\textbf {\bibinfo
  {volume} {12}},\ \bibinfo {pages} {114801} (\bibinfo {year}
  {2009})}\BibitemShut {NoStop}%
\bibitem [{\citenamefont {Hoover}(2024{\natexlab{a}})}]{hoover_four_2024}%
  \BibitemOpen
  \bibfield  {author} {\bibinfo {author} {\bibfnamefont {A.}~\bibnamefont
  {Hoover}},\ }\href@noop {} {\bibfield  {journal} {\bibinfo  {journal}
  {Physical Review Accelerators and Beams}\ }\textbf {\bibinfo {volume} {27}},\
  \bibinfo {pages} {122802} (\bibinfo {year} {2024}{\natexlab{a}})}\BibitemShut
  {NoStop}%
\bibitem [{\citenamefont {Lund}\ \emph {et~al.}(2006)\citenamefont {Lund},
  \citenamefont {Chilton},\ and\ \citenamefont {Lee}}]{lund_match_2006}%
  \BibitemOpen
  \bibfield  {author} {\bibinfo {author} {\bibfnamefont {S.~M.}\ \bibnamefont
  {Lund}}, \bibinfo {author} {\bibfnamefont {S.~H.}\ \bibnamefont {Chilton}}, \
  and\ \bibinfo {author} {\bibfnamefont {E.~P.}\ \bibnamefont {Lee}},\ }\href
  {\doibase 10.1103/PhysRevSTAB.9.064201} {\bibfield  {journal} {\bibinfo
  {journal} {Phys. Rev. ST Accel. Beams}\ }\textbf {\bibinfo {volume} {9}},\
  \bibinfo {pages} {064201} (\bibinfo {year} {2006})}\BibitemShut {NoStop}%
\bibitem [{\citenamefont {Hoover}\ \emph {et~al.}(2021)\citenamefont {Hoover},
  \citenamefont {Evans},\ and\ \citenamefont {Holmes}}]{hoover_matched_2021}%
  \BibitemOpen
  \bibfield  {author} {\bibinfo {author} {\bibfnamefont {A.}~\bibnamefont
  {Hoover}}, \bibinfo {author} {\bibfnamefont {N.~J.}\ \bibnamefont {Evans}}, \
  and\ \bibinfo {author} {\bibfnamefont {J.~A.}\ \bibnamefont {Holmes}},\
  }\href {\doibase 10.1103/PhysRevAccelBeams.24.044201} {\bibfield  {journal}
  {\bibinfo  {journal} {Phys. Rev. Accel. Beams}\ }\textbf {\bibinfo {volume}
  {24}},\ \bibinfo {pages} {044201} (\bibinfo {year} {2021})}\BibitemShut
  {NoStop}%
\bibitem [{\citenamefont {Baartman}(1998)}]{baartman_betatron_1998}%
  \BibitemOpen
  \bibfield  {author} {\bibinfo {author} {\bibfnamefont {R.}~\bibnamefont
  {Baartman}},\ }\href
  {https://www.osti.gov/etdeweb/servlets/purl/382303#page=282} {\bibfield
  {journal} {\bibinfo  {journal} {Proceedings of Space Charge Physics in High
  Intensity Hadron Rings (Shelter Island, New York, USA, 1998)}\ } (\bibinfo
  {year} {1998})}\BibitemShut {NoStop}%
\bibitem [{\citenamefont {Hoffman}\ \emph {et~al.}(1983)\citenamefont
  {Hoffman}, \citenamefont {Laslett}, \citenamefont {Smith},\ and\
  \citenamefont {Haber}}]{hoffman_stability_1983}%
  \BibitemOpen
  \bibfield  {author} {\bibinfo {author} {\bibfnamefont {I.}~\bibnamefont
  {Hoffman}}, \bibinfo {author} {\bibfnamefont {L.}~\bibnamefont {Laslett}},
  \bibinfo {author} {\bibfnamefont {L.}~\bibnamefont {Smith}}, \ and\ \bibinfo
  {author} {\bibfnamefont {I.}~\bibnamefont {Haber}},\ }\href@noop {}
  {\bibfield  {journal} {\bibinfo  {journal} {Particle Accelerators}\ }\textbf
  {\bibinfo {volume} {13}},\ \bibinfo {pages} {145} (\bibinfo {year}
  {1983})}\BibitemShut {NoStop}%
\bibitem [{\citenamefont {Evans}\ \emph {et~al.}(2024)\citenamefont {Evans},
  \citenamefont {Gorlov}, \citenamefont {Hoover},\ and\ \citenamefont
  {Morozov}}]{evans:hb2023}%
  \BibitemOpen
  \bibfield  {author} {\bibinfo {author} {\bibfnamefont {N.}~\bibnamefont
  {Evans}}, \bibinfo {author} {\bibfnamefont {T.}~\bibnamefont {Gorlov}},
  \bibinfo {author} {\bibfnamefont {A.}~\bibnamefont {Hoover}}, \ and\ \bibinfo
  {author} {\bibfnamefont {V.}~\bibnamefont {Morozov}},\ }in\ \href {\doibase
  10.18429/JACoW-HB2023-WEC3I1} {{\emph {\bibinfo
  {booktitle} {Proc. 68th Adv. Beam Dyn. Workshop High-Intensity
  High-Brightness Hadron Beams (HB'23)}}}},\ \bibinfo {series and number}
  {\bibinfo {series} {ICFA Advanced Beam Dynamics Workshop on High-Intensity
  and High-Brightness Hadron Beams}\ No.~\bibinfo {number} {68}}\ (\bibinfo
  {publisher} {JACoW Publishing, Geneva, Switzerland},\ \bibinfo {year}
  {2024})\ pp.\ \bibinfo {pages} {258--263}\BibitemShut {NoStop}%
\bibitem [{\citenamefont
  {Pelaia~II}(2016)}]{pelaia2016parameterestimationgaussiandampedsinusoids}%
  \BibitemOpen
  \bibfield  {author} {\bibinfo {author} {\bibfnamefont {T.}~\bibnamefont
  {Pelaia~II}},\ }\href {https://arxiv.org/abs/1604.05167} {\enquote {\bibinfo
  {title} {Parameter estimation of gaussian-damped sinusoids from a geometric
  perspective},}\ } (\bibinfo {year} {2016}),\ \Eprint
  {http://arxiv.org/abs/1604.05167} {arXiv:1604.05167 [physics.acc-ph]}
  \BibitemShut {NoStop}%
\bibitem [{\citenamefont {Schmidt}\ \emph {et~al.}(2002)\citenamefont
  {Schmidt}, \citenamefont {Forest},\ and\ \citenamefont
  {McIntosh}}]{Schmidt:573082}%
  \BibitemOpen
  \bibfield  {author} {\bibinfo {author} {\bibfnamefont {F.}~\bibnamefont
  {Schmidt}}, \bibinfo {author} {\bibfnamefont {E.}~\bibnamefont {Forest}}, \
  and\ \bibinfo {author} {\bibfnamefont {E.}~\bibnamefont {McIntosh}},\ }\href
  {https://cds.cern.ch/record/573082} {\emph {\bibinfo {title} {{Introduction
  to the polymorphic tracking code}}}},\ \bibinfo {type} {Tech. Rep.}\
  (\bibinfo  {institution} {CERN},\ \bibinfo {address} {Geneva},\ \bibinfo
  {year} {2002})\BibitemShut {NoStop}%
\bibitem [{\citenamefont {Grote}\ and\ \citenamefont
  {Schmidt}(2003)}]{grote2003mad}%
  \BibitemOpen
  \bibfield  {author} {\bibinfo {author} {\bibfnamefont {H.}~\bibnamefont
  {Grote}}\ and\ \bibinfo {author} {\bibfnamefont {F.}~\bibnamefont
  {Schmidt}},\ }in\ \href@noop {} {\emph {\bibinfo {booktitle} {Proceedings of
  the 2003 Particle Accelerator Conference}}},\ Vol.~\bibinfo {volume} {5}\
  (\bibinfo {organization} {IEEE},\ \bibinfo {year} {2003})\ pp.\ \bibinfo
  {pages} {3497--3499}\BibitemShut {NoStop}%
\bibitem [{mad()}]{madweb}%
  \BibitemOpen
  \href@noop {} {}\bibinfo {note}
  {https://madx.web.cern.ch/}\BibitemShut {NoStop}%
\bibitem [{ope()}]{openxal}%
  \BibitemOpen
  \href@noop {} {}\bibinfo {note}
  {https://openxal.github.io/}\BibitemShut {NoStop}%
\bibitem [{\citenamefont {Holmes}\ \emph
  {et~al.}(2018{\natexlab{b}})\citenamefont {Holmes}, \citenamefont {Gorlov},
  \citenamefont {Evans}, \citenamefont {Plum},\ and\ \citenamefont
  {Cousineau}}]{Holmes-Injection}%
  \BibitemOpen
  \bibfield  {author} {\bibinfo {author} {\bibfnamefont {J.~A.}\ \bibnamefont
  {Holmes}}, \bibinfo {author} {\bibfnamefont {T.}~\bibnamefont {Gorlov}},
  \bibinfo {author} {\bibfnamefont {N.~J.}\ \bibnamefont {Evans}}, \bibinfo
  {author} {\bibfnamefont {M.}~\bibnamefont {Plum}}, \ and\ \bibinfo {author}
  {\bibfnamefont {S.}~\bibnamefont {Cousineau}},\ }\href {\doibase
  10.1103/PhysRevAccelBeams.21.124403} {\bibfield  {journal} {\bibinfo
  {journal} {Phys. Rev. Accel. Beams}\ }\textbf {\bibinfo {volume} {21}},\
  \bibinfo {pages} {124403} (\bibinfo {year} {2018}{\natexlab{b}})}\BibitemShut
  {NoStop}%
\bibitem [{\citenamefont {Minerbo}(1979)}]{Minerbo}%
  \BibitemOpen
  \bibfield  {author} {\bibinfo {author} {\bibfnamefont {G.}~\bibnamefont
  {Minerbo}},\ }\href@noop {} {\bibfield  {journal} {\bibinfo  {journal}
  {Computer Graphics and Image Processing}\ }\textbf {\bibinfo {volume} {10}},\
  \bibinfo {pages} {48} (\bibinfo {year} {1979})}\BibitemShut {NoStop}%
\bibitem [{\citenamefont {Hoover}(2024{\natexlab{b}})}]{Hoover-4d}%
  \BibitemOpen
  \bibfield  {author} {\bibinfo {author} {\bibfnamefont {A.}~\bibnamefont
  {Hoover}},\ }\href {\doibase 10.1103/PhysRevAccelBeams.27.122802} {\bibfield
  {journal} {\bibinfo  {journal} {Phys. Rev. Accel. Beams}\ }\textbf {\bibinfo
  {volume} {27}},\ \bibinfo {pages} {122802} (\bibinfo {year}
  {2024}{\natexlab{b}})}\BibitemShut {NoStop}%
\bibitem [{\citenamefont {Shishlo}\ \emph {et~al.}(2015)\citenamefont
  {Shishlo}, \citenamefont {Cousineau}, \citenamefont {Holmes},\ and\
  \citenamefont {Gorlov}}]{PYROBIT}%
  \BibitemOpen
  \bibfield  {author} {\bibinfo {author} {\bibfnamefont {A.}~\bibnamefont
  {Shishlo}}, \bibinfo {author} {\bibfnamefont {S.}~\bibnamefont {Cousineau}},
  \bibinfo {author} {\bibfnamefont {J.}~\bibnamefont {Holmes}}, \ and\ \bibinfo
  {author} {\bibfnamefont {T.}~\bibnamefont {Gorlov}},\ }\href {\doibase
  https://doi.org/10.1016/j.procs.2015.05.312} {\bibfield  {journal} {\bibinfo
  {journal} {Procedia Computer Science}\ }\textbf {\bibinfo {volume} {51}},\
  \bibinfo {pages} {1272} (\bibinfo {year} {2015})},\ \bibinfo {note}
  {international Conference On Computational Science, ICCS 2015}\BibitemShut
  {NoStop}%
\bibitem [{\citenamefont {Burov}\ and\ \citenamefont
  {Derbenev}(2009)}]{burovDerbenev2009}%
  \BibitemOpen
  \bibfield  {author} {\bibinfo {author} {\bibfnamefont {A.}~\bibnamefont
  {Burov}}\ and\ \bibinfo {author} {\bibfnamefont {Y.}~\bibnamefont
  {Derbenev}},\ }\href {https://www.osti.gov/biblio/971378} {\emph {\bibinfo
  {title} {Space Charge Suppression for Uneven Emittances}}},\ \bibinfo {type}
  {Tech. Rep.}\ \bibinfo {number} {FERMILAB-PUB-09-392-AD}\ (\bibinfo {year}
  {2009})\BibitemShut {NoStop}%
\bibitem [{\citenamefont {Evans}\ \emph {et~al.}(2025)\citenamefont {Evans},
  \citenamefont {Hoover},\ and\ \citenamefont {Morozov}}]{dataset}%
  \BibitemOpen
  \bibfield  {author} {\bibinfo {author} {\bibfnamefont {N.}~\bibnamefont
  {Evans}}, \bibinfo {author} {\bibfnamefont {A.~M.}\ \bibnamefont {Hoover}}, \
  and\ \bibinfo {author} {\bibfnamefont {V.~Z.}\ \bibnamefont {Morozov}},\
  }\href {https://doi.org/10.5281/zenodo.15349989} {\enquote {\bibinfo {title} {{Data
  for Demonstration of a method to construct self-consistent beam distributions
  through phase space painting}},}\ } (\bibinfo {year} {2025}),\ \bibinfo
  {note} {{10.5281/zenodo.15349989}}\BibitemShut {NoStop}%
\end{thebibliography}

\end{document}